\documentclass[longauth]{aa} 
\usepackage{graphicx,natbib}
\begin{document}
   \title{Multifrequency variability of the blazar \object{AO 0235+164}\thanks{For questions
   regarding the availability of the data presented in this paper,
   please contact the WEBT President Massimo Villata ({\tt villata@to.astro.it})}}

   \subtitle{The WEBT campaign in 2004--2005 and long-term SED analysis}

   \author{C.~M.~Raiteri         \inst{ 1}
   \and   M.~Villata             \inst{ 1}
   \and   M.~Kadler              \inst{ 2,3}
   \and   M.~A.~Ibrahimov        \inst{ 4}
   \and   O.~M.~Kurtanidze       \inst{ 5}
   \and   V.~M.~Larionov         \inst{ 6}
   \and   M.~Tornikoski          \inst{ 7}
   \and   P.~Boltwood            \inst{ 8}
   \and   C.-U.~Lee              \inst{ 9}
   \and   M.~F.~Aller            \inst{10}
   \and   G.~E.~Romero           \inst{11,12}
   \and   H~.D.~Aller            \inst{10}
   \and   A.~T.~Araudo           \inst{11,12}
   \and   A.~A.~Arkharov         \inst{13}
   \and   U.~Bach                \inst{ 1}
   \and   D.~Barnaby             \inst{14}
   \and   A.~Berdyugin           \inst{15}
   \and   C.~S.~Buemi            \inst{16}
   \and   M.~T.~Carini           \inst{14}
   \and   D.~Carosati            \inst{17}
   \and   S.~A.~Cellone          \inst{12}
   \and   R.~Cool                \inst{18}
   \and   M.~Dolci               \inst{19}
   \and   N.~V.~Efimova          \inst{13}
   \and   L.~Fuhrmann            \inst{ 1,20}
   \and   V.A.~Hagen-Thorn       \inst{ 6}
   \and   M.~Holcomb             \inst{14}
   \and   I.~Ilyin               \inst{21}
   \and   V.~Impellizzeri        \inst{ 2}
   \and   R.~Z.~Ivanidze         \inst{ 5}
   \and   B.~Z.~Kapanadze        \inst{ 5}
   \and   J.~Kerp                \inst{22}
   \and   T.~S.~Konstantinova    \inst{ 6}
   \and   Y.~Y.~Kovalev          \inst{23,24,2}
   \and   Yu.~A.~Kovalev         \inst{24}
   \and   A.~Kraus               \inst{ 2}
   \and   T.~P.~Krichbaum        \inst{ 2}
   \and   A.~L\"ahteenm\"aki     \inst{ 7}
   \and   L.~Lanteri             \inst{ 1}
   \and   P.~Leto                \inst{25}
   \and   E.~Lindfors            \inst{15}
   \and   N.~Napoleone           \inst{26}
   \and   M.~G.~Nikolashvili     \inst{ 5}
   \and   K.~Nilsson             \inst{15}
   \and   J.~Ohlert              \inst{27}
   \and   I.~E.~Papadakis        \inst{28,29}
   \and   M.~Pasanen             \inst{15}
   \and   C.~Poteet              \inst{14}
   \and   T.~Pursimo             \inst{30}
   \and   E.~Ros                 \inst{ 2}
   \and   L.~A.~Sigua            \inst{31}
   \and   S.~Smith               \inst{14}
   \and   L.~O.~Takalo           \inst{15}
   \and   C.~Trigilio            \inst{16}
   \and   M.~Tr\"oller           \inst{ 7}
   \and   G.~Umana               \inst{16}
   \and   H.~Ungerechts          \inst{32}
   \and   R.~Walters             \inst{14}
   \and   A.~Witzel              \inst{ 2}
   \and   E.~Xilouris            \inst{33}
 }

   \offprints{C.~M.~Raiteri, \email{raiteri@to.astro.it}}

   \institute{
          INAF, Osservatorio Astronomico di Torino, 10025 Pino Torinese, Italy
   \and   Max-Planck-Institut f\"ur Radioastronomie, 53121 Bonn, Germany
   \and   NASA/Goddard Space Flight Center, Code 662, Greenbelt, Maryland 20771, USA
   \and   Ulugh Beg Astron.\ Inst., Academy of Sciences of Uzbekistan, Tashkent 700052, Uzbekistan
   \and   Abastumani Observatory, 383762 Abastumani, Georgia
   \and   Astron.\ Inst., St.-Petersburg State Univ., 198504 St.-Petersburg, Russia
   \and   Mets\"ahovi Radio Observatory, Helsinki Univ.\ of Technology, 02540 Kylm\"al\"a, Finland
   \and   1655 Stittsville Main St., Stittsville, Ont., Canada  K2S 1N6
   \and   Korea Astronomy and Space Science Institute, Republic of Korea
   \and   Dept.\ of Astronomy, Univ.\ of Michigan, Ann Arbor, MI 48109, USA
   \and   Inst.\ Argentino de Radioastronom\'{\i}a, 1894 Villa Elisa, Argentina
   \and   Facultad de Ciencias Astron\'omicas y Geof\'{\i}sicas, UNLP, La Plata, Argentina
   \and   Pulkovo Observatory, St.\ Petersburg, Russia                                                        
   \and   Dept.\ of Physics \& Astronomy, Western Kentucky Univ., Bowling Green, KY 42104, USA
   \and   Tuorla Observatory, FIN-21500 Piikki\"{o}, Finland                                                  
   \and   INAF, Osservatorio Astrofisico di Catania, 95123 Catania, Italy
   \and   Armenzano Astronomical Observatory, Assisi, Italy
   \and   Steward Observatory, 933 N.Cherry Ave. Tucson, AZ 85721, USA
   \and   INAF, Osservatorio Astronomico di Collurania Teramo, 64100 Teramo, Italy                            
   \and   Dipartimento di Fisica e Osservatorio Astronomico, Universit\`a di Perugia, Italy
   \and   Astrophysikalisches Institut Potsdam, D-14482 Potsdam, Germany
   \and   Argelander Institut f\"ur Astronomie, Universit\"at Bonn, D-53121 Bonn, Germany
   \and   National Radio Astronomy Observatory, Green Bank, WV 24944, USA
   \and   Astro Space Center of Lebedev Physical Inst., 117997 Moscow, Russia
   \and   INAF, Istituto di Radioastronomia, Sezione di Noto, 96017 Noto, Italy
   \and   INAF, Osservatorio Astronomico di Roma, 00040 Monte Porzio Catone, Italy
   \and   Michael Adrian Observatory, 65468 Trebur, Germany
   \and   IESL, FORTH, 711 10 Heraklion, Crete, Greece
   \and   Physics Dept., Univ.\ of Crete, 710 03 Heraklion, Crete, Greece
   \and   Nordic Optical Telescope, 38700 Santa Cruz de La Palma, Spain
   \and   Akhaltsikhe branch of the Tbilisi State University, Georgia
   \and   IRAM, Avd.\ Div.\ Pastora 7NC, 18012 Granada, Spain
   \and   Inst.\ of Astronomy and Astrophysics, National Observatory of Athens, 11810 Athens, Greece
 }

   \date{Received ...; accepted...}


  \abstract
   {}
   {A huge multiwavelength campaign targeting the blazar \object{AO 0235+164}
   was organized by the Whole Earth Blazar Telescope (WEBT) in 2003--2005 to study the
   variability properties of the source.}
   {Monitoring observations were carried out at cm and mm wavelengths, and in the near-IR and optical bands, 
   while three pointings by the XMM-Newton satellite provided information on the X-ray and UV emission.}
   {We present the data acquired during the second observing season, 2004--2005,
   by 27 radio-to-optical telescopes. The $\sim 2600$ data points collected allow us to trace the
   low-energy behaviour of the source in detail, revealing an increased near-IR and
   optical activity with respect to the previous season. Increased variability is also found
   at the higher radio frequencies, down to $\sim 15$ GHz, but not at the lower ones.
   While the X-ray (and optical) light curves obtained during the XMM-Newton pointings 
   reveal no significant short-term variability, the simultaneous intraday radio observations 
   with the 100 m telescope at Effelsberg show flux-density changes at 10.5 GHz,
   which are more likely due to a combination of intrinsic and extrinsic processes.}
   {The radio (and optical) outburst predicted to peak around February--March 2004 on the basis of
   the previously observed 5--6 yr quasi-periodicity did not occur.
   The analysis of the optical light curves reveals now a longer characteristic time scale of variability
   of $\sim 8$ yr, which is also present in the radio data.
   The spectral energy distributions corresponding to the XMM-Newton observations performed
   during the WEBT campaign are compared with those pertaining to previous pointings
   of X-ray satellites. Bright, soft X-ray spectra can be described in terms of an extra
   component, which appears also when the source is faint through a hard UV spectrum and a curvature
   of the X-ray spectrum. Finally, there might be a correlation between the X-ray and optical bright states
   with a long time delay of about 5 yr, which would require a geometrical interpretation.}
   
   \keywords{galaxies: active -- galaxies: BL Lacertae objects:
    general -- galaxies: BL Lacertae objects: individual:
    \object{AO 0235+164} -- galaxies: jets -- galaxies: quasars: general}

   \maketitle
%

\section{Introduction}

Among the active galactic nuclei, blazars (i.e.\ flat-spectrum radio quasars and
BL Lac objects) exhibit the most extreme properties: violent variability at all wavelengths,
high optical (and radio) polarization, superluminal motion of radio components, 
and brightness temperatures often exceeding
the inverse Compton limit.
These features are explained by assuming that the emission comes from
a relativistic plasma jet closely aligned with the line of sight \citep[see e.g.][]{bla78,bla79}.

One of the most intriguing blazars is AO 0235+164 (J0238+1636) at $z=0.94$ which,
in addition to the above characteristics, also shows a complex environment,
where a couple of foreground galaxies at $z=0.524$, possibly interacting,
can both significantly contaminate its emission
in the optical--UV band and absorb its radiation from the near-IR to the soft X-rays.
Moreover, the stars of these objects might act as microlenses, producing at least part
of the observed variability \citep[e.g.][]{ost85,sti88}.

The analysis of the radio and optical light curves of AO 0235+164 from 1975 to 2000 revealed
correlated variability and a quasi-periodicity of the main radio (and optical) outbursts
on a $\sim 5.7$ yr time scale \citep{rai01}.
The latter feature made this object a good candidate for
the binary black hole scenario \citep{rom03,ost04}, as in the case of OJ 287, another
famous blazar \citep[e.g.][]{sil88,vil98}.

In 2003--2005 AO 0235+164 has been the target of a huge multiwavelength observing effort
chiefly aimed at verifying its possible quasi-periodicity, since a radio outburst had
been predicted to peak around February--March 2004 (with an uncertainty of 6 months).
The Whole Earth Blazar Telescope\footnote{{\tt http://www.to.astro.it/blazars/webt/};
see e.g.\ \citet{vil04b,boe05,rai05}.}
(WEBT) consortium organized a radio, near-IR, and optical long-term monitoring campaign,
during which three observations by the XMM-Newton satellite were performed, with simultaneous
dense radio monitoring at the 100 m Effelsberg telescope.
A spectroscopic monitoring was also carried out at the VLT, TNG, and NTT telescopes
to study the relationship between the \ion{Mg}{ii} line flux and the continuum flux density,
and 15 VLBA epochs were granted to study the source structure variability.
The results of these observations will be presented elsewhere.

The results of the first observing season 2003--2004 were reported by \citet{rai05},
while a detailed analysis of the X-ray observations was performed by \citet{rai06a}.
Here we present the radio-to-optical data acquired by 27 telescopes
during the second observing season, 2004--2005.
The data of this extensive WEBT campaign 2003--2005, added to the historical light curves,
allow one to study the behaviour of this source over the last 30 yr.

This paper is organized as follows: the optical, near-IR, and radio light curves obtained during
the 2004--2005 observing season are presented in Sect.\ 2.
In Sect.\ 3 we derive optical colour indices over a longer time interval, including the 1989--1991
and 1997--1998
outbursts, to study the spectral variability of the source.
The intraday observations performed with the Effelsberg radio telescope during the August 2004
and January 2005 XMM-Newton pointings are presented and discussed in Sect.\ 4.
The long-term optical and radio behaviour of AO 0235+164 is presented in Sect.\ 5.
In Sect.\ 6 we analyse 11 SEDs of the source corresponding to different epochs of X-ray
observations, trying to understand how many emission components are involved.
Finally, the conclusions are drawn in Sect.\ 7.


\section{Observations during the 2004--2005 season}

The second part of the WEBT campaign took place from May 1, 2004 ($\rm JD = 2453127$)
to May 1, 2005 ($\rm JD = 2453492$). Table \ref{obs} shows the list of the participating
observatories, with the size of the telescope, the frequency/wavelength/band, 
and the number of observations performed.
As one can see, 2593 data points were acquired from 27 telescopes: 6 of them working at
cm wavelengths, one at mm wavelengths, one in the near-IR bands, and 19 in the 
optical bands.

   \begin{figure*}
   \centering
   \includegraphics{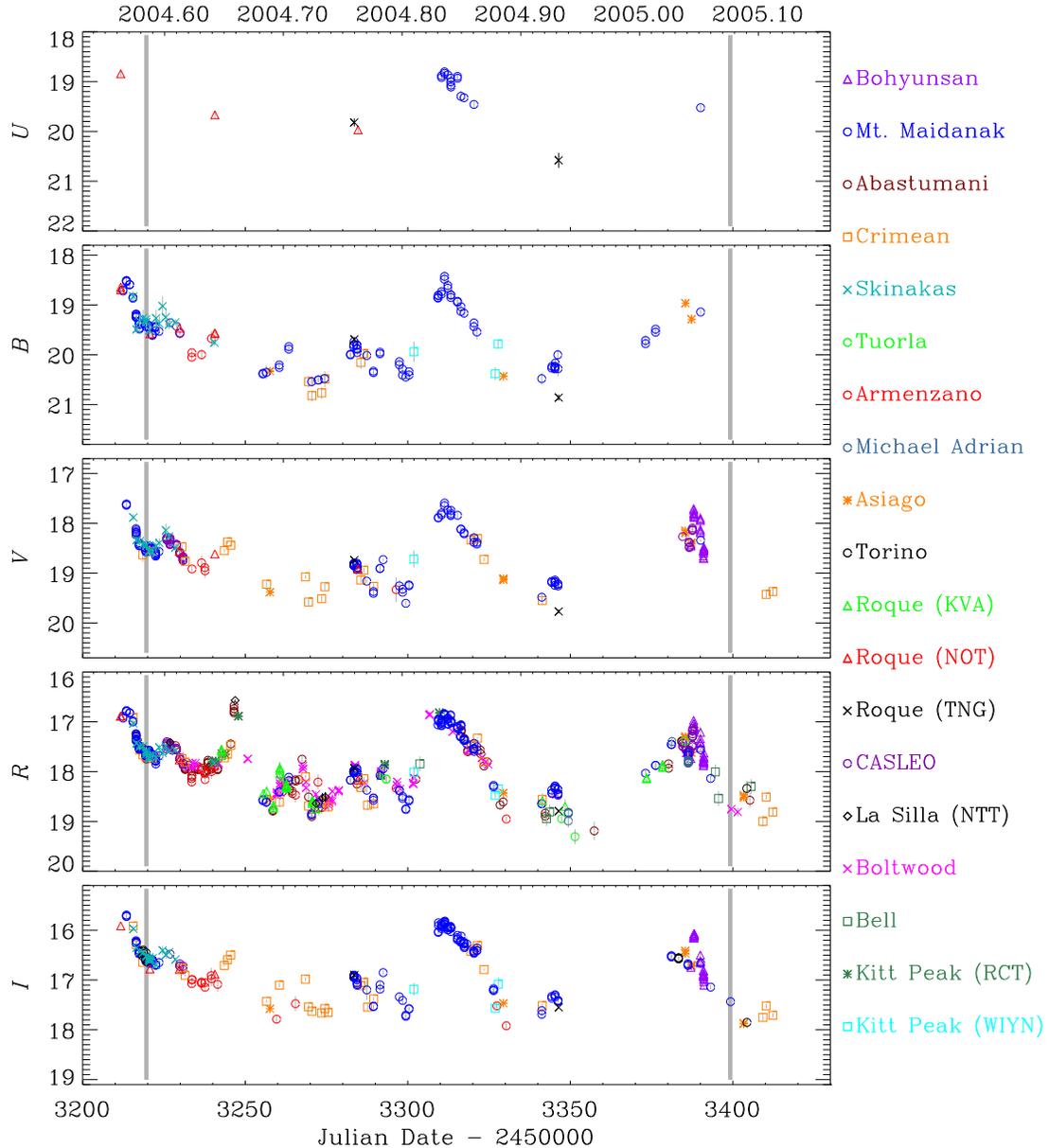}
      \caption{$UBVRI$ light curves of AO 0235+164 during the observing season 2004--2005;
      vertical lines indicate the XMM-Newton pointings of August 2, 2004 and January 28, 2005.
      The ELISA contribution has been subtracted as explained in \citet{rai05}.}
   \label{ubvri}
    \end{figure*}

\begin{table}
\caption{Ground-based observatories participating in the 2004--2005 observing season
of the WEBT campaign.}
\label{obs}
\centering
\begin{tabular}{l r c r }
\hline\hline
Observatory    & Tel. size     & Bands                & $N_{\rm obs}$\\
\hline
SAO RAS (RATAN)& 600 m$^a$     & 2.3, 4.8, 7.7,      & 33  \\
               &               & 11.1, 21.7 GHz      &     \\
Mets\"ahovi    & 14 m          & 37 GHz              & 142 \\
Noto           & 32 m          & 5, 8, 22 GHz        &  56 \\
Medicina       & 32 m          & 5, 8 GHz            &  23 \\
Effelsberg     & 100 m         & 4.9, 10.5 GHz       &  68 \\
UMRAO          & 26 m          & 4.8, 8.0, 14.5 GHz  & 112 \\
Pico Veleta    & 30 m          & 1, 2, 3 mm          &  18 \\
Campo Imperatore & 110 cm      & $J, H, K$           & 154 \\
Bohyunsan      & 180 cm        & $V, R, I$           & 126 \\
Mt. Maidanak   & 150 cm        & $U, B, V, R, I$     & 733 \\
Abastumani     &  70 cm        & $R$                 & 328 \\
Crimean        &  70 cm        & $B, V, R, I$        & 100 \\
Skinakas       & 130 cm        & $B, V, R, I$        &  83 \\
Tuorla         & 103 cm        & $R$                 &  20 \\
Armenzano      &  40 cm        & $B, V, R, I$        &  87 \\
Michael Adrian & 120 cm        & $R$                 &  48 \\
Asiago         & 182 cm        & $B, V, R, I$        &  38 \\
Torino         & 105 cm        & $R, I$              &  20 \\
Roque (KVA)    &  35 cm        & $R$                 &  69 \\
Roque (NOT)    & 256 cm        & $U, B, V, R, I$     &  24 \\
Roque (TNG)    & 358 cm        & $U, B, V, R, I$     &  10 \\
CASLEO         & 215 cm        & $B, V, R, I$        & 108 \\
La Silla (NTT) & 358 cm        & $R$                 &   5 \\
Boltwood       &  40 cm        & $R$                 & 129 \\
Bell           &  60 cm        & $R$                 &  23 \\
Kitt Peak (RCT)& 130 cm        & $R$                 &  24 \\
Kitt Peak (WIYN)& 90 cm        & $B, V, R, I$        &  12 \\
\hline
Total          &               &                     & 2593\\
\hline
\multicolumn{4}{l}{$^a$ Ring telescope}
\end{tabular}
\end{table}

Since the source was often very faint, a number of data were affected by large errors.
Hence, when constructing the light curves some data from the same telescope
were binned\footnote{With binning interval not exceeding half an hour.}
to increase the signal to noise ratio, while some other data,
in disagreement with the trend of the light curves, were discarded.
This cleaning process was mainly performed on the optical light curves, where
the comparison among the source behaviour in different bands and the better sampling
can help to recognize unreliable points.

Moreover, since the presence of an AGN only 2 arcsec south of the source
(named ELISA by \citealt{rai05}) affects the optical
photometry of AO 0235+164 especially in the blue part of the spectrum, 
we subtracted its contribution as discussed in \citet{rai05}.

\subsection{Optical and near-IR light curves}
   
Optical and near-IR data were collected as instrumental magnitudes of the source and of 
the reference stars in its field \citep{smi85,fio98} in order to apply the same magnitude
calibration to all datasets, when possible.

The $UBVRI$ light curves of the 2004--2005 observing season are shown
in Fig.\ \ref{ubvri}. Different symbols and colours are used to
distinguish the provenance of the data, as described in the legend.
Vertical lines are drawn through the dates of the XMM-Newton pointings of August 2, 2004
and January 28, 2005.

With respect to the previous observing season 
(see \citealt{rai05} and the top panel of Fig.\ \ref{radop}),
in 2004--2005 the source showed
increased variability, spanning more than 2.5 mag, although the average level
has remained rather faint.
In the $R$-band light curve, the best sampled one, four major flares can be recognized
(around $\rm JD = 2453210, 2453250, 2453310$, and 2453390),
which brought the source to be brighter than $R=17$.
The most dramatic variations were observed during the rising and dimming phases of the second flare,
when the $R$-band brightness first increased by $\sim 1.0$ mag in 2 days
and then decreased by $\sim 2.2$ mag in 12 days.

   \begin{figure}
   \centering
   \includegraphics[width=8cm]{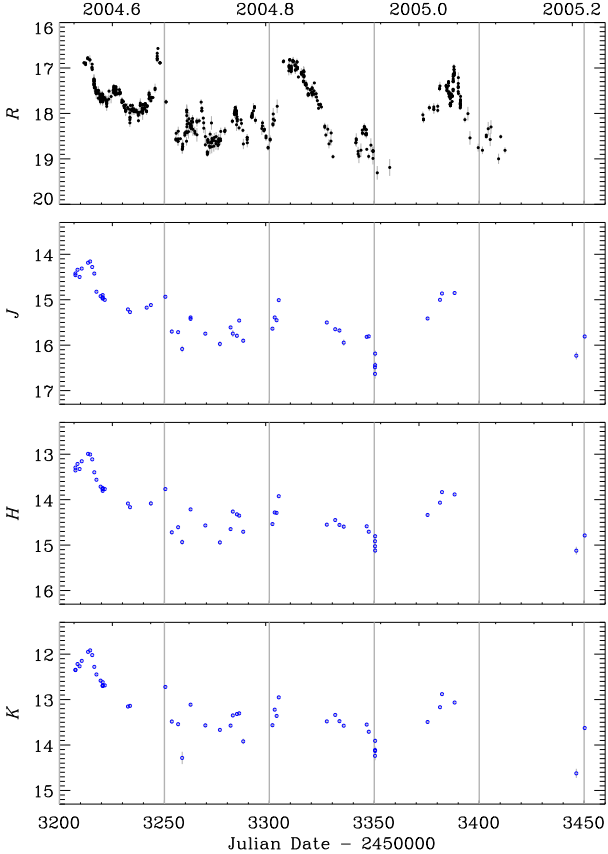}
      \caption{$JHK$ light curves of AO 0235+164 during the observing season 2004--2005 compared
      to the $R$-band one (top). Vertical lines are drawn to guide the eye 
      through variations in different bands.}
   \label{rjhk}
   \end{figure}

Near-IR monitoring was performed at the Campo Imperatore Observatory in the $JHK$ bands.
The resulting light curves are plotted in Fig.\ \ref{rjhk}, compared to the $R$-band one.
Although the near-IR light curves are less sampled than the $R$-band one,
a general agreement is visible. In particular, it is possible to recognize the first flare
around $\rm JD=2453210$ and the dip at $\rm JD=2453350$.

\subsection{Radio light curves}

A comparison between the $R$-band light curve and the radio behaviour at different
wavelengths is displayed in Fig.\ \ref{radop} for both
the observing seasons 2003--2004 and 2004--2005 of the WEBT campaign.
Data from the VLA/VLBA Polarization Calibration Database are
also shown\footnote{\tt http://www.vla.nrao.edu/astro/calib/polar/}.
The vertical lines mark the XMM-Newton pointings.
Cubic splines through the 30-day binned light curve at 37 GHz and
50-day binned light curves at 22, 14.5, 8, and 5 GHz are drawn to
highlight the long-term variations\footnote{We chose a shorter time bin for the
37 GHz data because the light curve at 37 GHz is better sampled than those at the lower frequencies 
and flux density variations are faster.}. Horizontal lines show the
magnitude and flux-density average values over the considered period: 
$<$$R$$>$ $=17.84$,
$<$$F_{37}$$>$ $=1.66 \, \rm Jy$,
$<$$F_{22}$$>$ $=1.32 \, \rm Jy$,
$<$$F_{14.5}$$>$ $=1.44 \, \rm Jy$,
$<$$F_{8}$$>$ $=1.42 \, \rm Jy$,
$<$$F_{5}$$>$ $=1.49 \, \rm Jy$.
These 2003--2005 optical and radio levels are relatively low if compared with those reached during
the major historical outbursts; indeed, when looking at the historical light curves 
(Fig.\ \ref{longterm}) one can see that
$R_{\rm min}=14.03$, 
$F_{37,\rm max}= 6.88 \, \rm Jy$,
$F_{22,\rm max}= 6.10 \, \rm Jy$, 
$F_{14.5,\rm max}= 8.09 \, \rm Jy$,
$F_{8,\rm max}= 7.03 \, \rm Jy$, and
$F_{5,\rm max}= 4.28 \, \rm Jy$.

As already found for AO 0235+164 and for other blazars, 
variations have larger flux amplitudes at the higher frequencies.
It is interesting to notice that a higher flux density is visible at 37 GHz
at the beginning of the 2004--2005 observing season, when the source was experiencing
an optical flare. The radio flux density from 37 to 14.5 GHz was also higher at the time
of the optical flare which occurred around $\rm JD = 2453300$--2453330,
while at the start of the 2003--2004 season some radio activity at all frequencies 
seems to follow the bright optical state.
However, it is hard to say whether these higher radio states have some correlation
with their contemporaneous optical events, or if they are delayed counterparts
of previous optical flares or, finally, if there is no correlation at all.
Actually, each of these possibilities has already been singled out in the historical behaviour
of the source \citep[e.g.][ and references therein]{rai05}.

   \begin{figure*}
   \centering
      \caption{$R$-band magnitudes (top) and radio fluxes (Jy) at different frequencies
      during the 2003--2005 WEBT campaign. Vertical lines mark the times of the three
      XMM-Newton pointings. Cubic splines on the radio data at 37, 22, 14.5, 8, and 5 GHz
      are plotted to better see the long-term trend. Horizontal lines indicate
      the mean fluxes at these frequencies. Two different symbols are used for the 37 GHz data
      from Mets\"ahovi since they come from different datasets.}
   \label{radop}
    \end{figure*}

\section{Optical colour indices}

The colour behaviour of blazars can be an important tool to disentangle different
components contributing to the optical emission variability.
In the case of BL Lacertae, \citet{vil02,vil04a} showed that the short-term variations are strongly
chromatic, with a bluer-when-brighter behaviour, likely due to intrinsic energetic processes.
On the contrary, the longer-term flux changes do not imply important spectral variations, 
and are possibly of geometric origin.
A recent study of the quasar-type blazar 3C 454.3 \citep{vil06} shows that the optical 
spectrum steepens
when the source is in outburst and the jet contribution to the optical flux is dominant.
During fainter optical states instead, the emission from a blue thermal disc may be responsible
for the observed spectral flattening (bluer-when-fainter behaviour).
The multi-colour optical observations of 8 blazars (3 quasars and 5 BL Lacs) by \citet{gu06} support this result,
showing that quasars tend to be redder when they are brighter, while BL Lacs tend to be bluer.
An analysis of the spectral behaviour of a sample of blazars was performed by \citet{fio04}:
in particular, they found that the mean spectral slope is flatter during the rising phases of flares,
while it is steeper during the declines.

\begin{figure*}
   \centering
   \includegraphics{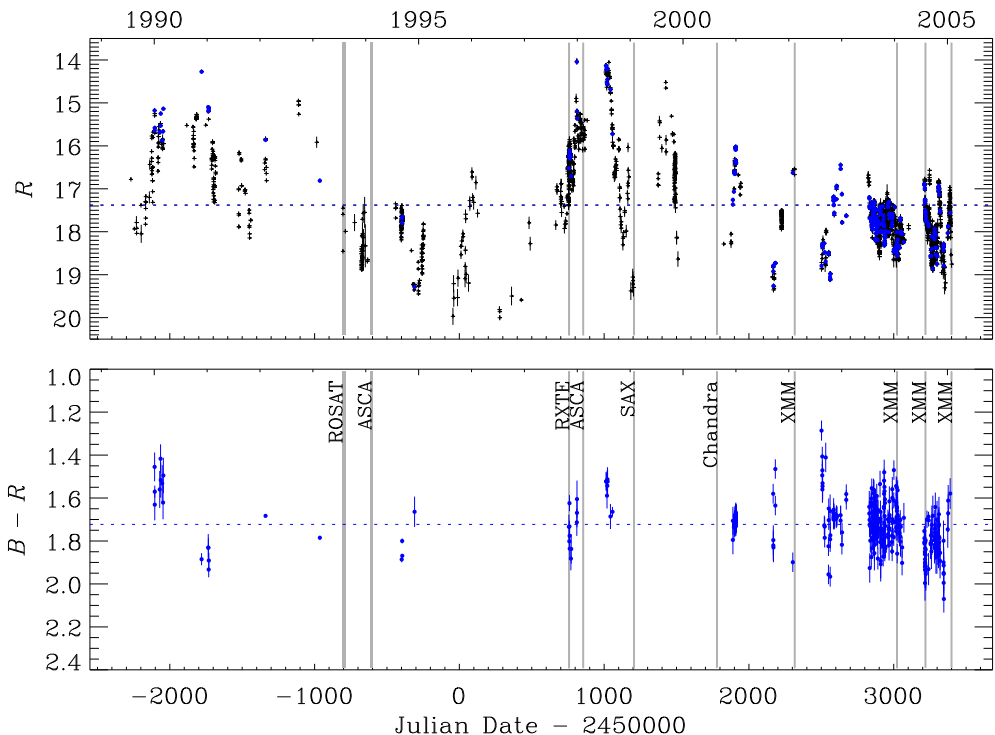}
      \caption{$B-R$ colour indices (bottom) compared to the $R$-band light curve (top);
      in the top panel blue (larger) dots indicate the $R$ data used to derive the colour indices.
      Horizontal lines are drawn through average values; vertical lines indicate
      the times of the X-ray satellite pointings, discussed in Sect.\ 5.}
   \label{colori}
   \end{figure*}

We show the $B-R$ colour index of AO 0235+164 as a function of time in Fig.\ \ref{colori} (bottom), 
compared to the $R$-band light curve (top). The plot covers the period 1989--2005 to include the last
major optical outbursts. The $B-R$ points have been obtained by coupling each $B$ magnitude from a given telescope
with the average of the $R$ data from the same telescope acquired next the $B$
datum (i.e.\ within 10--20 minutes). Only measurements with errors not greater than 0.08 mag were considered.
In this way, we coupled 401 $B$ data with 670 $R$ ones to obtain indices with a maximum error of 0.10 mag.

The colour index shows a noticeable scatter, from 1.27 to 2.07, which implies important spectral variations.
The average value is 1.73, with a standard deviation of 0.11.
However, it is not possible to recognize any simple correlation with brightness.
In particular, when considering the indices corresponding to the outburst phases of
1989--1991 and 1997--1998, they do not show peculiar values, since similar $B-R$ values 
are found also in much fainter states.
Thus, it seems that the long-term behaviour of the source is
essentially achromatic. The $V-I$ colour indices confirm this finding.

However, one can notice a progressive spectral reddening in the last years: although
the average $R$ magnitude\footnote{Obtained
by considering only the points which are used to calculate the colour indices.}
in the 2002--2003, 2003--2004, and 2004--2005 observing seasons is similar ($\sim 17.8$),
the average $B-R$ values are 1.66, 1.72, and 1.82, respectively.

Finally, one can derive the optical spectral index $\alpha_{\rm opt}$ ($F_\nu \propto \nu^{-\alpha}$)
from the colour index:
$\alpha_{\rm opt} = [(B-R)-(A_B-A_R)-0.3521]/0.4125$.
By setting the extinction values $A_B$ and $A_R$ equal to 1.904 and 1.260, respectively (see Table 5
in \citealt{rai05}),
we obtain an average value $\alpha_{\rm opt}=1.76$, with standard deviation 0.26.
Although this value has to be taken with some caution because it is calculated from two bands only,
it nonetheless gives an indication of the optical spectral behaviour of the source.
The spectrum appears rather steep, but still within the range of spectral indices of low-energy peaked
BL Lacs derived by \citet{fio04}.
The very high values of $\alpha_{\rm opt}$ ($>2$) estimated for AO 0235+164 by
various authors in previous works were obtained by neglecting the absorption
due to the $z=0.524$ intervening system.

\section{Intraday radio observations at Effelsberg}

  \begin{figure}
   \centering
   \includegraphics[width=8cm]{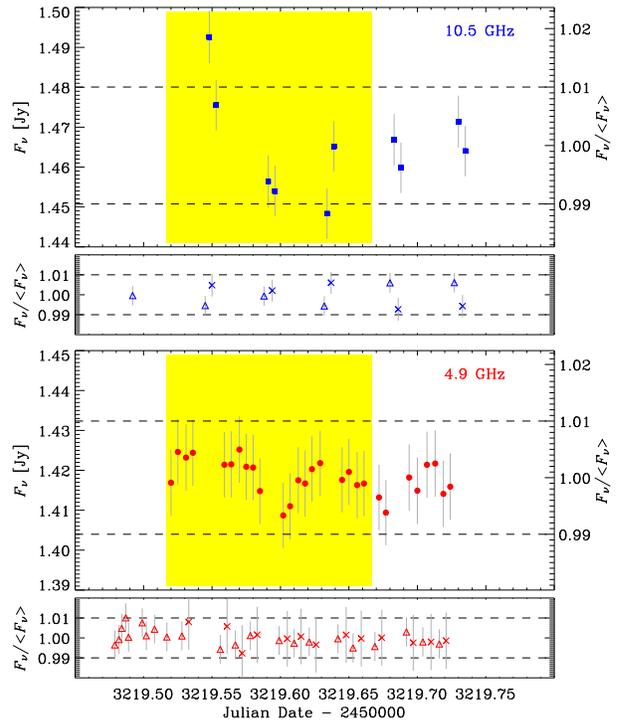}
      \caption{Radio light curves of the source and calibrators
      at 10.5 GHz (top, blue symbols) and 4.9 GHz (bottom, red symbols)
      obtained with the 100 m radio telescope at Effelsberg on August 2, 2004.
      The two calibrators are 3C 067 (triangles) and 4C +09.11 (crosses).
      The horizontal lines indicate deviations of 1\% from the mean flux densities,
      while the yellow areas highlight the period of the X-ray observations.}
   \label{effe2}
   \end{figure}
   \begin{figure}
   \centering
   \includegraphics[width=8cm]{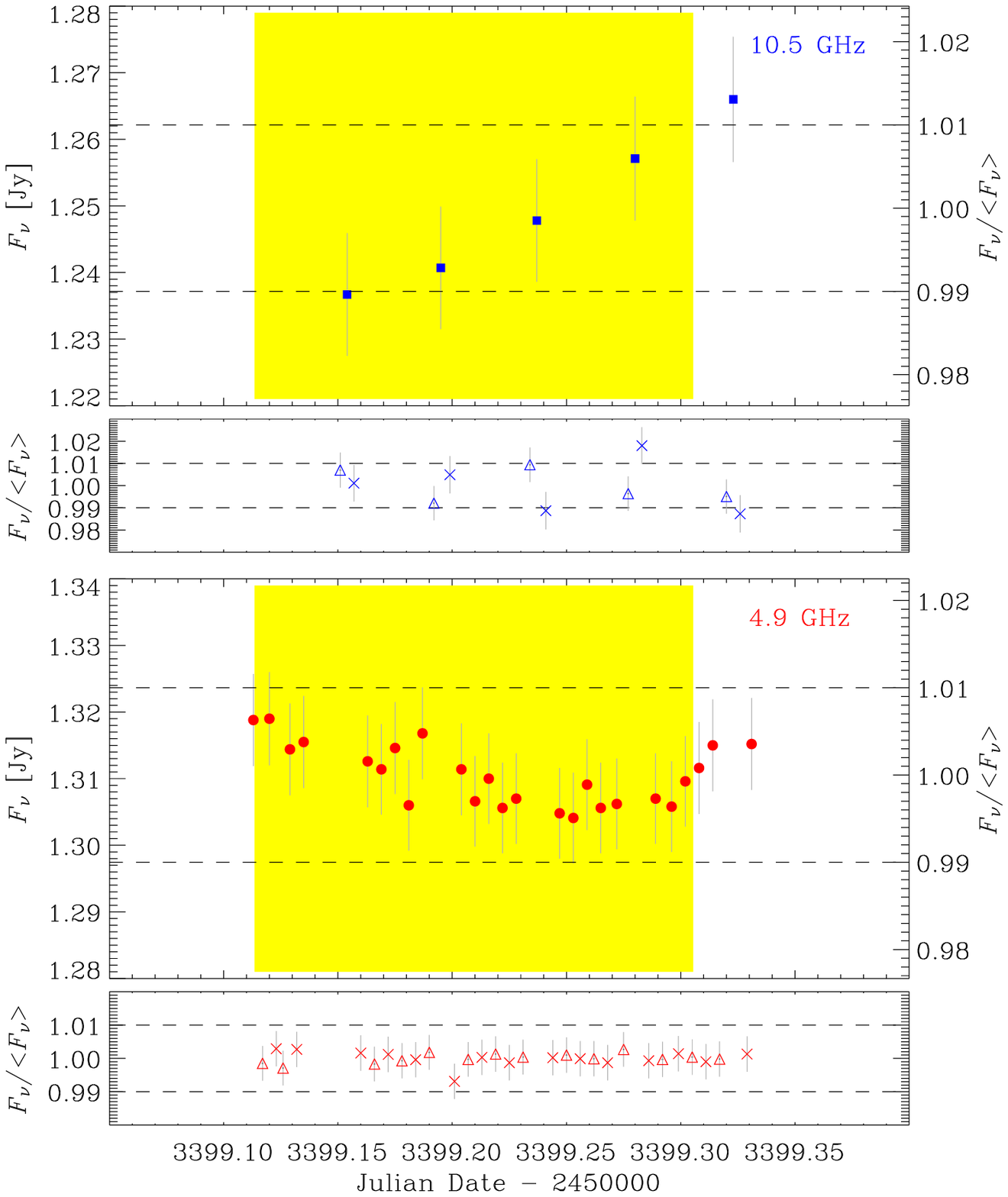}
      \caption{Radio light curves of the source and calibrators
      at 10.5 GHz (top, blue symbols) and 4.9 GHz (bottom, red symbols)
      obtained with the 100 m radio telescope at Effelsberg on January 28, 2005.
      The two calibrators are 3C 067 (triangles) and 4C +09.11 (crosses).
      The horizontal lines indicate deviations of 1\% from the mean flux densities,
      while the yellow areas highlight the period of the X-ray observations.}
   \label{effe3}
   \end{figure}
As in the case of the XMM-Newton pointing of January 18--19, 2004 discussed
in \citet{rai05}, also during the other two pointings of August 2, 2004 and
January 28, 2005, densely sampled radio observations were performed at 4.9 and 10.5 GHz
with the 100 m radio telescope at Effelsberg.
The observing strategy and data reduction procedures are based on \citet{kra03}
and have been outlined in \citet{rai05}.
The resulting light curves are shown in Fig.\ \ref{effe2} for the August 2004 pointing
and in Fig.\ \ref{effe3} for the January 2005 one.
The yellow areas highlight the period of the X-ray observations\footnote{The X-ray light curves
corresponding to the three XMM-Newton pointings have been analysed in \citet{rai06a}; they show no
significant variability. Further analysis of the radio and X-ray light curves can be found in \citet{kad05}.}.
Normalized flux densities of the two calibrators 3C 067 (J0224+2750) (triangles) 
and 4C +09.11 (J0238+1010) (crosses) are shown in the bottom panels.
Horizontal lines in both the source and the calibrators panels indicate 1\% deviations from the mean 
flux densities. They allow to grab the amount of variability and to compare the source one
with that of the calibrators.

In both epochs the 4.9 GHz light curve does not show significant
variability, while the flux density changes at 10.5 GHz are more interesting.
Indeed, on August 2, 2004 one can see first a sharp flux decrease of 3\% 
in about 2 hours\footnote{Calculated as the variation between the first (the highest)
and fifth (the lowest) point, which has also a slope equal to that determined by the second--fourth
points. A similar, but more uncertain, slope is also simultaneously present in the 4.9 GHz data.},
and then a slower flux increase.
On January 28, 2005 the percent variation of the source flux is not greater than the calibrators one,
but the fact that we see a quasi-linear trend suggests that this can be a piece of
a larger variation, and that the probability that it is noise is very low.

Table \ref{stat} reports some statistics:
in both epochs the standard deviation $\sigma$ is greater than the root mean square uncertainty
$\epsilon=\sqrt{\sum_{i=1}^{N}{\epsilon_i ^2}/N}$ (where $\epsilon_i$ are the individual errors)
at 10.5 GHz, while it is smaller at 4.9 GHz.
The same conclusion is reached by considering the reduced chi-square $\chi^2_{\rm r}$, which
tests the hypothesis that the light curve can be modelled by a constant function:
the values in Table \ref{stat} tell that this hypothesis is fair for the 4.9 GHz data,
but not for the 10.5 GHz ones.
Finally, Table \ref{stat} also shows the modulation index $m \, [\%] = 100 \, \sigma /$$<$$F_\nu$$>$, which
provides a measure of the strength of the observed variation amplitudes without
taking into account the errors of the individual measurements.
This is lower at 4.9 GHz than at 10.5 GHz and inferior to
what was found for the January 18--19, 2004 light curves, when the modulation index was
1.2\% at 10.5 GHz and 0.5\% at 4.9 GHz.

\begin{table}
\caption{Statistics on the Effelsberg data}
\label{stat}
\centering
\begin{tabular}{l c c c c c}
\hline\hline
Freq. & $<$$F_\nu$$>$ & $\sigma$ & $\epsilon$ & $\chi^2_{\rm r}$ & $m$ \\
(GHz) & (Jy)          & (Jy)     & (Jy)       &                  & (\%) \\
\hline
\multicolumn{6}{c}{August 2, 2004}\\
10.5  & 1.465     & 0.013    & 0.006 & 3.76 & 0.89\\
4.9   & 1.418     & 0.004    & 0.008 & 0.29 & 0.28\\
\hline
\multicolumn{6}{c}{January 28, 2005}\\
10.5  & 1.250     & 0.012    & 0.009 & 1.66 & 0.96\\
4.9   & 1.311     & 0.005    & 0.007 & 0.45 & 0.38\\
\hline
\end{tabular}
\end{table}

The limited duration of the radio monitoring allows neither to fully constrain the true time scale
of the 10.5 GHz variations nor to check whether flux changes at 4.9 GHz led or lagged those
detected at the higher frequency. Hence, we are not able to say whether these flux changes are of
intrinsic nature or if interstellar scattering is responsible for at least part of the variations.
Also a higher modulation index at higher frequency may be reconciled with interstellar scintillation
by invoking a strong scintillation regime \citep[see e.g.][]{ric90}.

However, in the following we speculate about the physical implications of these variations
in case we ascribed them to an intrinsic process.

Under the assumption of an intrinsic nature of the 10.5 GHz flux density variations
detected in August 2, 2004,
the brightness temperature corresponding to the initial fast decrease, 
calculated following \citet{wag96}\footnote{We adopt
a flat cosmology with $H_0=71 \, \rm km \, s^{-1} \, Mpc^{-1}$ and $\Omega=0.27$,
which yields a source luminosity distance of $\sim 6140$ Mpc.},
is $T_{\rm b} \sim 7 \times 10^{17} \, \rm K$, and a Doppler factor $\delta > 87$
would be required to bring the rest frame value below the inverse Compton limit
of $10^{12} \, \rm K$ ($T_{\rm b,observed}= \delta^3 T_{\rm b,intrinsic}$), 
confirming previous findings of a very high Doppler factor to explain
some observing features of this source \citep{kra99,fuj99,fre00,jor01}.
If we take the data errors into account, the inferred brightness temperature can actually
vary from $\sim 3$ to $\sim 11 \times 10^{17} \, \rm K$, corresponding to
lower limits to $\delta$ of 70 and 104, respectively.

On January 28 the 10.5 GHz light curve shows a quasi-linear increasing trend, characterized
by a $\sim 2$\% flux variation in about 4 hours. 
Under the assumption of intrinsic variability, the associated brightness temperature
would be $T_{\rm b}\sim 9 \times 10^{16} \, \rm K$, implying $\delta > 45$.
This value of the Doppler factor is similar to that estimated by \citet{rai05} from the
Effelsberg observations of January 2004 ($\delta > 46$).
Although it is much smaller than that derived from the August 2004 observations,
nonetheless it would imply strong relativistic beaming of the emitted radiation
in the intrinsic variability scenario. When considering the data errors, the lower limit 
to $\delta$ can actually range between 23 and 62 for this variation.

We have to mention that
recent results, obtained during the current relatively quiescent state of the source, point to
$\delta$ and $T_{\rm b}$ values lower than the extreme values
estimated in the above mentioned earlier papers.
\citet{fre06} analysed VSOP data acquired in 2001--2002.
They found a decrease of the Doppler boosting by a factor of $\sim 20$ with respect to the value
$\delta \sim 100$ inferred from data taken in 1999 just after a major outburst \citep{fre00}.
Their results are consistent with those derived by \citet{kov05} from VLBA 2-cm Survey data
of March 2001.
Apparent speeds up to $(25.6 \pm 7)c$ were measured by \citet{pin06} in the jet of AO 0235+164
based on VLBA observations during 2002--2003. These values are quite smaller than those previously 
reported by \citet{jor01} and suggest Doppler factors not exceeding $\sim 50$.

This boosting decrease can be explained, for instance, by an increased viewing angle of the
radio emitting region in the
helical jet scenario proposed by \citet{vil99} and applied to the case of AO 0235+164 by \citet{ost04}.
In this case, the high brightness temperature value we inferred from the August 2 data 
would be overestimated and the observed variation would be due, at
least in part, to interstellar scattering.

It is interesting to notice that the brightness temperature derived from VLBI
observations performed on September 11, 2004 at 15 GHz is only $2.6 \times 10^{11} \, \rm K$
(MOJAVE collaboration, priv.\ comm.).
Although both the VLBI and IDV brightness temperatures
are lower limits, they are much more different than it is usually found for these compact objects
\citep[see e.g.][]{zen02}, favouring an extrinsic explanation for the flux density
variation observed on August 2, 2004.

    \begin{figure*}
      \centering
      \caption{Historical optical (mag)
      and radio (37, 22, 14.5, 8, and 5 GHz, flux density in Jy) light curves of AO 0235+164
      shifted by the indicated amount; to increase the sampling in the optical, we show $R$-band data
      after $\rm JD=2447000$ and $R$-converted $B$-band data before.
      Yellow strips highlight the periods (a, b, c, d, e, f)
      where a major radio (and optical) outburst is expected on the basis of a
      $5.6 \pm 0.5$ yr quasi-periodicity; periods $\alpha$, $\beta$, and $\gamma$
      belong to an $\sim 8$ yr characteristic time scale of variability.}
   \label{longterm}
   \end{figure*}

\section{Long-term behaviour}

The historical behaviour of the optical magnitudes and radio fluxes is plotted in Fig.\ \ref{longterm}.
These light curves have been shifted by a certain amount (indicated in the figure) to better distinguish
the trend at different wavelengths. To increase the sampling in the optical, we plotted $R$-band data after
$\rm JD=2447000$ and $B$-band data converted into $R$ ones before, according to
$R=B-$$<$$B-R$$>$, with $<$$B-R$$>$ $=1.7$.

The five yellow strips named a, b, c, d, and e
highlight the radio outbursts (with optical counterpart) which were found to repeat
quasi-periodically by \citet{rai01}.
Indeed, the 1 yr thick strips, equally spaced by 5.6 yr, contain the peak of the
most prominent radio (and optical) outbursts.
The last strip (f) indicates the period when the next outburst should have occurred according
to the \citet{rai01} prediction, which motivated the WEBT campaign.

It is evident that the last $\sim 5$ yr of suppressed activity
in the radio bands are very atypical for this source.
    
\begin{figure}
   \centering
   \includegraphics[width=8cm]{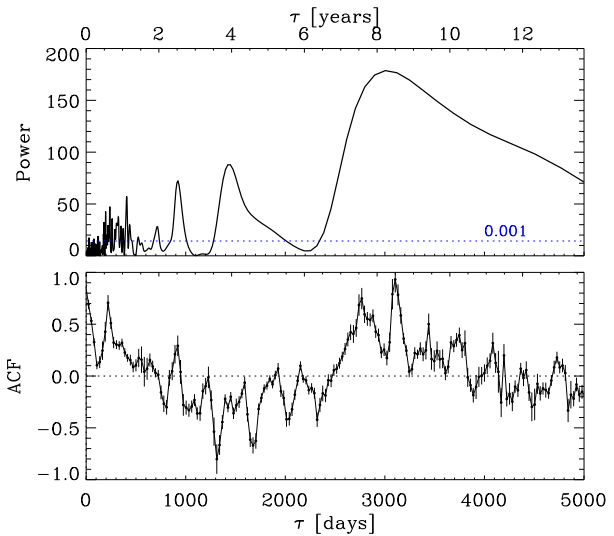}
      \caption{Power spectrum of the discrete Fourier transform (top) and
      autocorrelation function (bottom) obtained from the historical $R$-band light curve.
      In the top panel the blue dotted line indicates the level above which
      the significance is better than 0.001.}
   \label{auto_r}
   \end{figure}

\begin{figure}
   \centering
   \includegraphics[width=8cm]{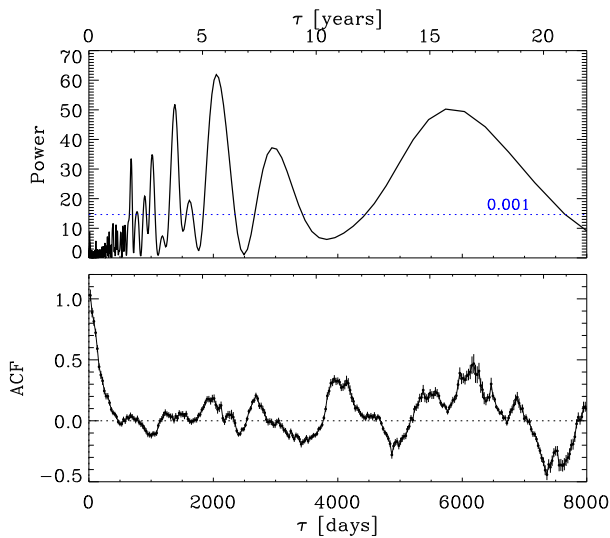}
      \caption{Power spectrum of the discrete Fourier transform (top) and
      autocorrelation function (bottom) obtained from the historical 8 GHz light curve.
      In the top panel the blue dotted line indicates the level above which the significance
      is better than 0.001.}
   \label{auto_8}
   \end{figure}

If we run a periodicity analysis on the historical optical light curves applying the
discrete Fourier transform \citep[DFT;][]{pre92},
the most significant signal occurs at $\sim 3000$ days (8.2 yr).
A similar time scale can be found now also in the radio light curves,
together with the 5--6 yr period
found by \citet{rai01}\footnote{We point out that the present analysis is performed on different
light curves with respect to \citet{rai01}: not only they are more extended in time, but also more
datasets have been added to both the optical and radio light curves to improve the sampling of the
pre-2000 period, as discussed in \citet{rai05}. In particular, besides the radio data from Mets\"ahovi
and UMRAO, we included data (both recent and past ones) from RATAN, Noto, Medicina, Effelsberg, and NRAO.}.
We show these results in Figs.\ \ref{auto_r} and \ref{auto_8}.

Figure \ref{auto_r} (top panel) displays the results of the DFT applied to the historical
$R$-band light curve. Before running the DFT algorithm,
data have been binned over 1 day intervals to homogenize the sampling.
The significance of peaks in the DFT, 
which correspond to sinusoidal components of period $\tau$ in the light curve,
is estimated by the false alarm probability, i.e.\ the probability
that a peak is higher than a certain level if the data are pure noise.
The horizontal (dotted blue) line in the figure marks the value of power corresponding 
to a significance level better than 0.001.
The DFT indicates that the most likely period is 8.2 yr,
with other possible periods at
3.9 and 2.5 yr. The numerous signals corresponding to time scales around (or less than)
1 yr may be spurious, since they may reflect gaps in the data train.
The bottom panel of Fig.\ \ref{auto_r} displays the
autocorrelation function \citep[ACF;][]{ede88,huf92,pet04} of the $R$-band data.
The most remarkable feature is that it shows two prominent peaks at $\sim 7.6$ and $\sim 8.5$ yr.
The reason is that this method, which is more sensitive to maxima and minima brightness levels
than to sinusoidal modulations of the light curves, feels the double-peaked nature of some events
and the differences in the peak distances (see Fig.\ \ref{longterm}).

A similar plot for the 8 GHz light curve, the most extended and best sampled among the radio datasets,
is presented in Fig.\ \ref{auto_8}. The DFT has been obtained after binning the data over 1 week, thus
making the sampling more homogeneous.
The strongest signal corresponds to the 5.6 yr period, 
which has not been suppressed by the lack of the predicted outburst in 2004, followed by
signals at 3.7, 15.7, 8.0, 2.8, 1.9 yr, and by minor ones.
We notice that some of the time scales are roughly multiples of the others, and that
in the ACF plot the strongest peaks are found at about 11 and 15--17 yr, i.e.\ at twice the
5.6 and 8 yr periods detected with the DFT. However, even for these peaks the value of the 
ACF does not exceed 0.5.

We also tested our results with the structure function \citep[SF;][]{sim85} method,
and found that its results are very similar to those already obtained with the ACF.

An $\sim 8$ yr characteristic time scale of variability
would put into correlation the outbursts occurred in the periods
labelled $\alpha$, $\beta$, and $\gamma$ in Fig.\ \ref{longterm}.
If this $\sim 8$ yr recurrence should repeat,
an outburst will be visible during the 2006--2007 observing season.

We also performed a cross-correlation analysis by means of the discrete correlation function
(DCF) on the historical $R$-band and radio light curves in order to check whether the addition
of a further observing season of data can modify the results reported by \citet{rai05}.
Both the time lags of the radio variations with respect to the optical ones
(from 1 to 2.5 months depending on the frequency separation) and
the delays among variations at different radio frequencies
(from 10 days to 1.5 months) found by \citet{rai05} are confirmed.

\section{Multiepoch spectral energy distribution}

\begin{figure*}
   \centering
   \includegraphics{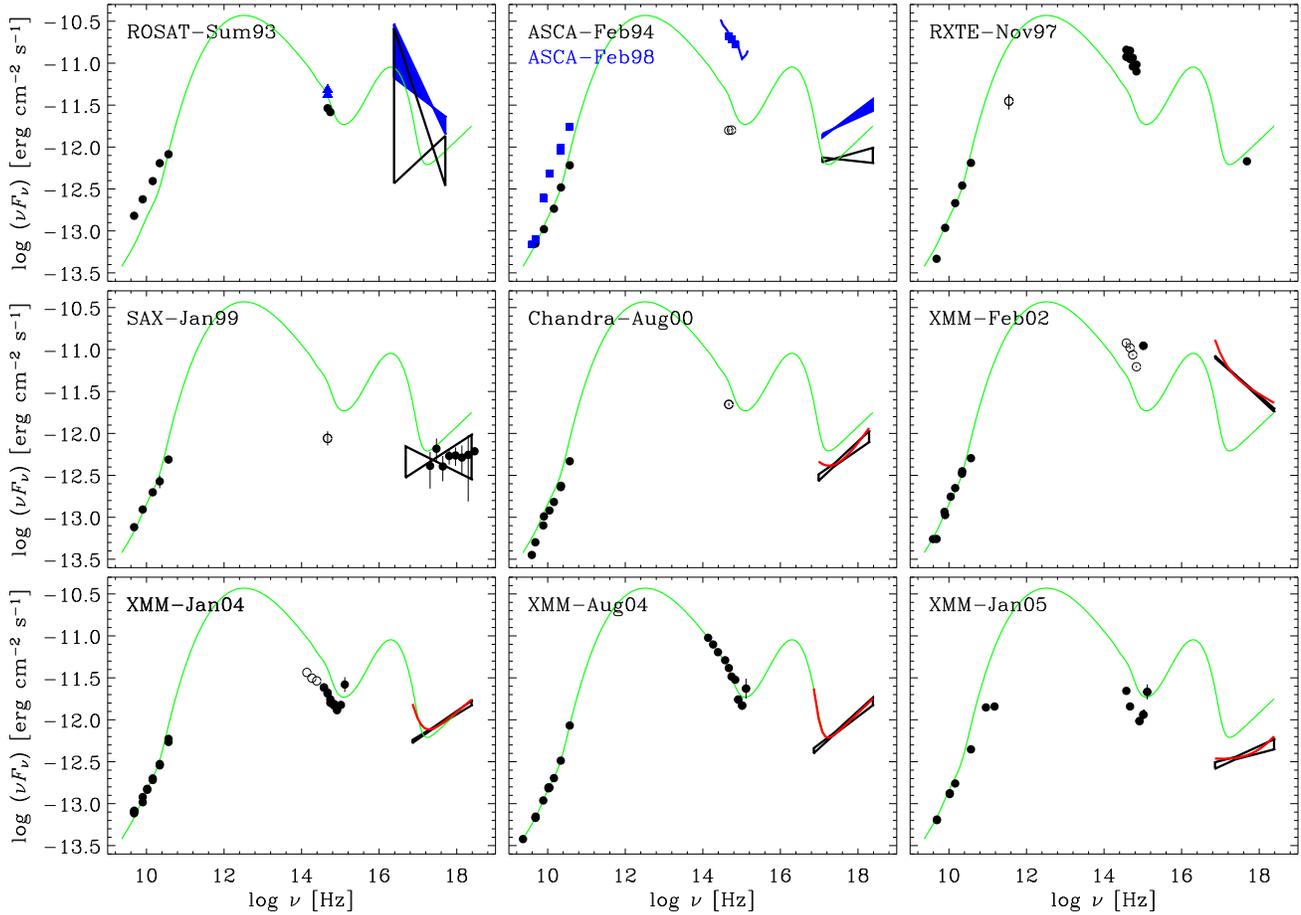}
      \caption{Spectral energy distributions of AO 0235+164 during X-ray
      satellite pointings. The green solid line represents a cubic spline
      interpolation through the August 2004 data.}
   \label{sed}
    \end{figure*}

The broad-band spectral energy distributions (SEDs) of AO 0235+164 corresponding
to the times when X-ray satellites
observed the source are displayed in Fig.\ \ref{sed}.
Only epochs after 1990 were considered, thus neglecting
the uncertain results obtained by Einstein in 1979--1981 and by EXOSAT in 1984.
An analysis of the XMM-Newton and Chandra observations can be found in \citet{rai06a}, where a summary of
previous X-ray observations by other satellites is also given together with the pertaining references.

For the Chandra and XMM-Newton X-ray spectra, we display the results of both single power law (black bow-tie)
and double power law (red line) models with Galactic absorption plus
absorption from the intervening system at $z=0.524$ obtained by \citet{rai06a}.
Those authors favoured the latter, curved solution.
For all epochs but the February 2002 one, for which the curved solution was by far statistically superior,
this was mainly motivated by the fact that in this energy region we expect the superposition
of the synchrotron and inverse-Compton components.
We also display the UV and optical fluxes in the $M2$, $W1$, $U$, $B$, and $V$ bands 
derived by \citet{rai06a} by analysing the observations of the Optical Monitor onboard XMM-Newton.

In the figure, optical (and near-IR) data simultaneous to the X-ray observations are plotted with filled symbols;
when simultaneous data were missing, we searched for the closest data available, and showed them with empty symbols.
Since radio variations are slower than the optical ones, we plotted as filled symbols both strictly simultaneous
radio data and data obtained as averages of those acquired just before and just after the satellite pointing.

Optical data were corrected for the ELISA contribution.
Near-IR and optical magnitudes were de-reddened according to the prescriptions given by \citet{jun04},
adopting the values tabulated by \citet{rai05}, and then transformed into fluxes with the zeropoint 
values given by \citet{bes98}.

Radio flux densities at 4.8, 8.0, and 14.5 GHz are from UMRAO,
those at 22 and 37 GHz are from the Mets\"ahovi Radio Observatory. 
RATAN simultaneous data at 0.9, 2.3, 3.9, 4.8, 7.7, 11.1, and 21.7 GHz contributed
to some SEDs (ASCA-Feb98, Chandra-Aug00, XMM-Feb02, XMM-Aug04). In the last three SEDs,
corresponding to the XMM-Newton pointings inside the WEBT campaign, 
radio data at 4.9 and 10.5 GHz from the Effelsberg telescope have been added.

The green solid line plotted in each panel represents a cubic spline interpolation through the
August 2004 data; it was drawn only to help the comparison of SEDs.

As one can see from Fig.\ \ref{sed}, the flat radio spectrum can always be connected with the steep optical
one through what is called ``the synchrotron bump". This is likely peaking in the far-infrared,
as low-energy peaked BL Lac objects are expected to do.
Most of the times the X-ray spectrum is hard,
suggesting that we are seeing the first part of the inverse-Compton bump, which is expected to peak 
in the GeV energy range. However, in summer 1993 and February 2002 the X-ray
spectrum is soft. As for the ROSAT spectrum, \citet{mad96} suggested that in a bright state
the peak of the synchrotron component may shift to higher energies, and thus the synchrotron radiation
would dominate in the soft X-ray region. This interpretation was also followed by
\citet{pad04}, when fitting the ROSAT spectrum with the homogeneous, one-zone model of \citet{ghi02}.
However, this scenario is questioned by the contemporaneous optical data shown in Fig.\ \ref{sed}.
Indeed, by looking at the various SEDs of this figure, there are no clues that the peak frequency
may strongly change depending on the brightness state of the source.
Actually, the situation may be more complex, since
a UV--soft-X-ray bump, i.e.\ an extra SED component in addition to
the synchrotron and inverse-Compton ones, seems to be present at least in some cases,
as discussed by \citet{jun04} and \citet{rai05,rai06a}.

In the following, we briefly examine the individual SEDs, with particular attention to the optical
data, since the presence of optical data simultaneous with the X-ray ones is fundamental to understand
the emission mechanisms operating in the different energy bands.

\subsection{ROSAT: Summer 1993}
The ROSAT pointing lasted from July 21 to August 15, 1993.
The X-ray spectrum displayed as a blue, filled bow-tie in the figure refers to the analysis performed by
\citet{mad96} on the full dataset and reveals a bright and soft X-ray spectrum.
However, since the ROSAT light curve showed significant variability,
\citet{com97} reanalysed the data identifying two epochs of higher and lower count rates.
The first epoch (July) yielded results similar to those of \citet{mad96}.
$R$-band data taken on July 27 by the Hamburg Quasar Monitoring program\footnote{\tt
http:///www.friedensblitz.de/hqm/} are plotted in the figure
as blue triangles. The connection of these data with the X-ray spectrum seems to require an extra bump.

The second ROSAT epoch (August) produced a different spectrum, which is plotted in the figure as
a black, empty bow-tie.
It is simultaneous with the acquisition of the $V$ and $R$ data displayed as black dots, which were
taken at the Lowell Observatory on August 13.
In this case the extra bump is not strictly needed, but also not ruled out.

\subsection{ASCA: February 1994 and 1998}
The first ASCA pointing (black, empty bow-tie) lasted from February 4 to 19, 1994 \citep{mad96}.
The closest optical data (black circles) were acquired at the JKT (La Palma) on January 13 \citep{tak98}.

ASCA pointed at the source again on February 11--12, 1998, when it was in a bright radio and optical state
(see Figs.\ \ref{colori} and \ref{longterm}). 
However, the X-ray flux was not particularly high, and the X-ray spectrum was harder
(blue, filled bow-tie) than during the previous pointing \citep{jun04}.
A simultaneous optical--UV spectrum was acquired with the HST \citep{jun04}, which is displayed as a blue line in
Fig.\ \ref{sed}, and simultaneous $BVR$ data were taken with the REOSC telescope of the 
Torino Observatory (blue squares), well matching the HST spectrum.
The misalignment of the HST and X-ray spectra was discussed by \citet{jun04},
who considered the possibility of an extra SED component as unlikely and
suggested instead that it may arise from the difficulties in modelling the extinction of the $z=0.524$ absorber.

It is interesting to point out here the correlation between the radio and X-ray spectral states: the harder and
brighter X-ray spectrum of 1998 corresponds to a brighter and flatter radio spectrum, 
as expected when considering the inverse-Compton origin of the X-ray radiation.

\subsection{RXTE: November 1997}
The RXTE pointing was performed on November 3--6, 1997, during the rising phase of the optical outburst
(see Fig.\ \ref{colori}).
The X-ray flux was found low \citep{web00}.
Simultaneous optical data were taken by many observatories in the ambit of the first-light WEBT campaign \citep{rai01}.
In  Fig.\ \ref{sed} we plotted two $BVRI$ spectra obtained at the Teide Observatory on November 3 and 6 in order
to have homogeneous data. Also shown is a point at 850 $\mu$m obtained by averaging data taken
at the JCMT before and after the RXTE observations \citep{web00}.

\subsection{BeppoSAX: January 1999}
BeppoSAX observed AO 0235+164 on January 28, 1999. The bow-tie in Fig.\ \ref{sed} 
shows the results of the analysis by
\citet{pad04}, while the points in the X-ray domain were obtained directly by the 
ASI BeppoSAX Data Center\footnote{\tt http://www.asdc.asi.it/bepposax/}.
The $R$-band datum was taken at the Torino Observatory 5 days before.

\subsection{Chandra: August 2000}
The Chandra observations were performed on August 20--21, 2000. They revealed a faint
and hard X-ray spectrum \citep{tur03,rai06a}.
The closest optical datum was taken at the Torino Observatory 7 weeks later.

\subsection{XMM-Newton: February 2002}
The XMM-Newton observations of February 10, 2002 were analysed in detail by \citet{rai06a}
(see also \citealt{kad05,fos06}).
The source was found in the brightest X-ray state
ever detected ($F_{1 \, \rm keV} \sim 2 \, \mu \rm Jy$), with a soft spectrum, and the X-ray
light curve revealed noticeable variability on a time scale of $\sim 40$ minutes.
The Optical Monitor (OM) onboard XMM-Newton observed only in the UV$W1$ band,
finding a bright UV state.
Three $R$-band frames were acquired at the Mount Maidanak Observatory $\sim 1.5$ days before the XMM-Newton
observations, revealing a moderately bright state ($R \sim 16.6$).
In the figure we plotted a complete $BVRI$ optical spectrum taken at Mount Maidanak 
11 days earlier, when the source was exactly at the same optical level.
However, an $I$-band datum taken at the Torino Observatory 9 days after the XMM-Newton
pointing showed the source to be much fainter ($I \sim 16.9$), so it is possible that at the
time of the XMM-Newton observation the source was in a dimming phase, and that simultaneous optical
data would place just below those shown in the figure.
In conclusion, it is very unlikely that the optical, UV, and X-ray data
can be fitted by a single convex synchrotron component.
The UV--soft-X-ray bump would instead provide a comfortable way to describe this SED.

\subsection{XMM-Newton: January 2004}
The SED corresponding to the XMM-Newton pointing of January 18--19, 2004
was described in detail in \citet{rai05}.
The contemporaneous radio data come from Mets\"ahovi, Effelsberg, and UMRAO,
the optical (and near-IR) data were taken at the NOT, while the OM observed in the $V$, $B$, $U$,
UV$W1$, and UV$M2$ filters.
The transition from a steep optical spectrum to a flat UV one is dramatic, and
the probable curvature of the X-ray spectrum
supports the existence of the UV--soft-X-ray bump.

\subsection{XMM-Newton: August 2004}
The plotted radio points at 4.9 and 10.5 GHz are averages of the intraday observations at Effelsberg
(see Fig.\ \ref{effe2}); the 14.5 GHz datum is from UMRAO;
the 37 GHz point represents the average of the Mets\"ahovi data.
Data at 2.3, 4.8, 7.7, 11.1, and 21.7 GHz taken simultaneously by RATAN
at the beginning of September have been added
to increase the sampling. They nicely agree with the other radio data.
Near-IR information comes from Campo Imperatore.
Many optical data are available in the $BVRI$ bands, with a fair sampling, showing a constant level
without any significant variation.
We display the averages of the data taken
at Skinakas, to have a homogeneous $BVRI$ set. Other $BVRI$ sequences acquired at Mount Maidanak
just before the start of the XMM-Newton observations confirm the Skinakas average spectrum.
The hints for the extra bump come from the UV$M2$ datum and the strong
curvature of the X-ray spectrum when a double power law model is applied.

\subsection{XMM-Newton: January 2005}
As in the previous epoch, the points at 4.9 and 10.5 GHz represent the averages of the Effelsberg data
(see Fig.\ \ref{effe3}); the 14.5 GHz datum is from UMRAO, the 37 GHz one was obtained at Mets\"ahovi,
and the 2 and 3 mm data were acquired at Pico Veleta.
In the optical band, the $I$ datum is from Mount Maidanak and the $R$ datum is from Boltwood.
Here the curvature of the X-ray spectrum is smaller than in the previous epoch, but there are 
two UV points (UV$W1$ and UV$M2$ from the OM) that may support the existence of a UV bump.

\bigskip
In summary, the existence of the UV--soft-X-ray bump in the SED of AO 0235+164 is supported by two arguments:
the SED hardening in the UV domain, which gives rise to a concave optical--UV branch, and the corresponding
``rise" of the soft-X-ray spectra towards the UV band.
Even if the first argument should fall in case the UV extinction in the $z=0.524$ intervening system had been
strongly overestimated by \citet{jun04}, the extra-bump cannot be avoided in the cases where the X-ray spectrum
is soft.

\section{Discussion and conclusions}

Among blazars, AO 0235+164 presents several peculiar features. It has shown dramatic
flux changes at all wavelengths on a wide range of timescales, including IDV.
Its main outbursts have repeated quasi-periodically, at least for some time, 
with clear correlation of the radio and optical flux variations with very short time delays.
When considering this along with the presence of an intervening galactic system,
also gravitational microlensing becomes a viable interpretation for the source variability.
At milliarcsecond scales it shows an extremely compact core, with faint extensions occasionally
appearing at various position angles. Several observations suggest that the source radiation is 
unusually highly Doppler-boosted.

In this paper we presented the results of the radio, near-IR, and optical monitoring
during the second observing season of the WEBT campaign 2003--2005 targeting AO 0235+164. 
The results of the first observing season and a detailed analysis
of the three XMM-Newton observations performed during the campaign have already been
reported in \citet{rai05,rai06a}.

The big radio outburst, with a possible optical counterpart, which was
foreseen to peak around February--March 2004
on the basis of the quasi-periodicity of $5.7 \pm 0.5$ yr
discussed by \citet{rai01}, did not occur. 
The radio flux remained rather low during the whole WEBT campaign, with some mild
increased activity at the higher radio frequencies only in the last observing season 2004--2005.
A mild increased activity in the last season also characterized the optical and near-IR light curves.
The comparison of the optical and radio data taken during the WEBT campaign with
the long-term light curves of this object shows how this last, long period of suppressed activity
especially at radio wavelengths is unusual for this source.
The analysis of the historical light curves reveals now a
longer characteristic time scale of variability of $\sim 8$ yr,
more evident in the optical, but also present in the radio data 
along with the $\sim 5.7$ yr one.

The colour-index analysis reveals noticeable spectral changes in the optical band, which nonetheless
do not show correlation with the source magnitude.
The long-term trend appears essentially achromatic,
with a possible progressive reddening in the last years.

All the three XMM-Newton observations performed during the WEBT campaign found the source
in a faint state at all wavelengths.
Neither the optical nor the  X-ray light curves show any significant variability.
In contrast, the simultaneous intraday radio observations at the 100 m Effelsberg telescope
reveal some flux changes at 10.5 GHz. 
If they are ascribed to intrinsic processes, a Doppler factor up to $\sim 100$ can in principle be derived, 
confirming previous findings that this is an extremely beamed source. However, on the basis of these data
we cannot rule out that interstellar scintillation was responsible for at least part of the 
variability. Actually, a reduced Doppler factor would agree with recent findings
pointing to a reduced beaming during the current relatively quiescent period.

We compiled the SEDs corresponding to the X-ray observations since the ROSAT pointing in summer 1993.
Radio and optical data simultaneous with the X-ray observations have been looked for
in our database in order to shed light on the various emission components involved.
The aim was mainly to find clues to the presence of the UV--soft-X-ray bump already discussed
by \citet{jun04} and \citet{rai05,rai06a}.
The existence of this extra component, besides the classical synchrotron and inverse-Compton ones,
is mainly supported by the UV data, when available,
and by the soft X-ray spectra detected by ROSAT in summer 1993 and by XMM-Newton in February 2002.

A discussion on the possible interpretations of this bump can be found in \citet{rai06a}.
Among the various pictures, these authors focused mainly on the presence of a thermal accretion disc
and on an additional synchrotron component.
In the latter case, which was favoured by the authors, we can envisage two possibilities.
Either there are actually two synchrotron components, 
coming from distinct and separate jet regions without
any simple relationship, or we are in the presence of
a continuous inhomogeneous jet whose radiation is for
some reason suppressed in the optical--near-UV region. 
Here we have again two possibilities:
i) the break in the spectrum is due to a discontinuity in the opacity 
(as in the case of Mkn 501, \citealt{rai03p}),
ii) the suppressed emission is caused by a misalignment of the corresponding 
jet region whose radiation is less beamed towards the line of sight\footnote
{Although at other wavelengths (and hence spatial resolution), this scenario
has indeed been suggested by \citet{bac06} to interpret the region
of suppressed flux density in the intensity profiles of the VLBA images of
BL Lacertae.}.
The observed strong variability in both these SED components might favour a scenario
where both the emission from the near-IR to the soft X-rays
and the spectral gap change in time as the strongly bent jet precesses or rotates.
This view may be supported by the large variations of the VLBI jet components
position angles observed by many authors
\citep[see e.g.][]{fre00,fre06,jor01,pin06}.

We noticed how among the ten satellite pointings in 1993--2005,
only two of them (ROSAT in summer 1993 and XMM-Newton in February 2002)
found the source in a bright and soft X-ray state, possibly of synchrotron origin.
The All Sky Monitor X-ray light curve 
of AO 0235+164\footnote{\tt http://heasarc.nasa.gov/docs/xte/ASM/sources.html}
from January 1996 to November 2005
shows that X-ray bright states are indeed very uncommon for this source:
only in the first half of 2002 a significant increase of the count rate was observed, in correspondence
to the first XMM-Newton pointing.

The two bright X-ray states are separated by about 8.5 yr, and a $\sim 9$ yr interval
elapsed between the ROSAT and the 1984 EXOSAT observation,
when a high (but very uncertain) X-ray flux was measured.
A similar time separation is found when considering
the last strongest optical outbursts of 1990 and 1998. Thus, both of them
seem to follow an X-ray brightening by about 5 yr.
It will be interesting to see whether an optical flux increase will occur 
in the next observing seasons,
i.e.\ about 5 yr after the 2002 X-ray bright state.
It would support both the possible quasi-periodicity of $\sim 8$ yr of the optical
(and radio) outbursts and a correlation between the optical and X-ray emission with a long time delay.
A correlation of this type would be explained by a scenario where a curved inhomogeneous jet
precesses or rotates, thus subsequently aligning its different emitting portions with the line of sight.

Several WEBT members are continuing to monitor the source in the radio, near-IR, and optical bands
so that we will be able to check whether the source activity continues to increase and eventually leads to
a forthcoming outburst.

\begin{acknowledgements}
We thank the staff of the Asiago Observatory for performing the optical observations
of this paper in service mode and the staff of the Medicina Radio Station for assistance 
during the observations.
Based on observations with the 100 m telescope of
the MPIfR (Max-Planck-Institut f\"ur Radioastronomie) at Effelsberg.
Based on observations made with the Nordic Optical Telescope, operated
on the island of La Palma jointly by Denmark, Finland, Iceland,
Norway, and Sweden, in the Spanish Observatorio del Roque de los
Muchachos of the Instituto de Astrof\'{\i}sica de Canarias.
Based on observations made with the Italian Telescopio Nazionale Galileo (TNG)
operated on the island of La Palma by the Centro Galileo Galilei of the INAF
(Istituto Nazionale di Astrofisica) at the Spanish Observatorio del Roque de los Muchachos
of the Instituto de Astrof\'{\i}sica de Canarias.
Based on observations made at Complejo Astron\'omico El Leoncito, operated
under agreement among CONICET, UNLP, UNC, and UNSJ, Argentina.
This research has made use of data from the University of Michigan Radio Astronomy Observatory,
which is supported by the National Science Foundation and by funds from the University of Michigan.
\mbox{RATAN--600} observations were partly supported by the
Russian State Program ``Astronomy'' (project~1.2.5.1), the Russian
Ministry of Education and Science, and the Russian Foundation for Basic
Research (projects 01--02--16812, 02--02--16305, 05--02--17377).
This work was partly supported by the European Community's Human Potential Programme
under contract HPRN-CT-2002-00321 (ENIGMA) 
and by the Italian Space Agency under contract ASI-INAF I/023/05/0.
V.M.L. acknowledges support from Russian Federal Foundation for Basic Research (RFBR), 
project 05-02-17562.
MK has been supported in part by a Fellowship of the International Max Planck Research 
School for Radio and Infrared Astronomy at the Universities of Bonn and Cologne and in 
part by an appointment to the NASA Postdoctoral Program at the Goddard Space Flight Center, 
administered by Oak Ridge Associated Universities through a contract with NASA.
This project was done while Y.Y.K. was a Jansky fellow of the 
National Radio Astronomy Observatory and a research fellow of the 
Alexander von Humboldt Foundation.
\end{acknowledgements}

\end{document}